\documentclass[letter]{aa}

\usepackage{txfonts}
\usepackage{graphicx}
\usepackage{amsmath}
\usepackage{xcolor}
\usepackage{placeins}
\usepackage{ragged2e}
\usepackage{hyperref}
\usepackage{breakurl}
\usepackage{tikz}
\usetikzlibrary{shapes,arrows}

\begin{document}

\title{The evolution of the gas fraction of quiescent galaxies modeled as a consequence of 
their creation rate}

\author{R. Gobat\inst{1}
\and G. Magdis\inst{2,3,4}
\and C. D'Eugenio\inst{5}
\and F. Valentino\inst{2,4}
}

\institute{
Instituto de F\'{i}sica, Pontificia Universidad Cat\'{o}lica de Valpara\'{i}so, Casilla 4059, 
Valpara\'{i}so, Chile
\and Cosmic Dawn Center (DAWN), Copenhagen, Denmark. 
\and DTU-Space, Technical University of Denmark, Elektrovej 327, DK-2800 Kgs. Lyngby, Denmark
\and University of Copenhagen, Jagtvej 128, DK-2200 Copenhagen N, Denmark
\and CEA, Irfu, DAp, AIM, Universit\'{e} Paris-Saclay, Universit\'{e} de Paris, CNRS, F-91191 
Gif-sur-Yvette, France
}

\date{}

\abstract{
We discuss the evolution of the interstellar medium of quiescent galaxies, currently 
emerging from recent analyses, with the help of a simple model based on well-established 
empirical relations such as the stellar mass functions and the main sequence of star formation.
This model is meant to describe observed quantities without making specific assumptions on the 
nature of quenching processes, but relying on their observable consequences. 
We find that the high gas fractions seen or suggested at high redshift in quiescent galaxies, 
and their apparent mild evolution at early times, can be mostly attributed to a progenitor 
effect where recently quenched galaxies with $\sim$10\% gas fractions dominate the quiescent 
galaxy population until $z\sim1$. In the same context, the much lower gas and dust fractions 
measured in local early-type galaxies are interpreted as the product of the steady depletion 
of their interstellar medium on a $\sim2$\,Gyr timescale, coupled with a higher fraction of 
more gas-exhaustive events.}


\titlerunning{Modeling the gas fraction of QGs}
\authorrunning{Gobat et al.}

\maketitle

\section{\label{wot?}Introduction}

Early-type galaxies (ETGs) are the eventual end product of galaxy evolution and, as such, 
are overwhelmingly passive systems relative to the overall coeval galaxy population. 
Consequently, their faint interstellar medium (ISM) has historically received relatively 
little attention, at least compared to that of star-forming late-type galaxies, except in 
the local Universe where it is much more easily detectable. 
This changed recently, following the systematization of mid- and far-infrared (FIR) astronomy that 
allowed for (and was spurred by) an increased understanding of star formation (SF) in galaxies. 
The ISM of quiescent galaxies can indeed inform us about the mechanisms mediating the cessation 
(hereafter, ``quenching'') of star formation, since some processes, such as the suppression of gas 
infall and cooling \citep[e.g.,][]{BD03,Cro06} or the gravitational stabilization by galactic spheroids 
\citep{Mar09,Tac15},  are expected to leave untapped reservoirs in these galaxies. 
Interestingly, the currently emerging picture suggests that, while local ETGs are largely gas-poor 
\citep[e.g.,][]{You11}, their higher-redshift counterparts can have gas fractions 
($M_{\text{gas}}/(M_{\star}+M_{\text{gas}})$, hereafter $f_{\text{gas}}$) that are higher by a factor 10-100 
\citep[hereafter G18]{Rud17,Spi18,Gob18}. 
On the other hand, most studies so far have used samples that were selected ad hoc and, consequently, 
this inferred evolution does not account for progenitor bias. 
Here we present a simple model to interpret the observed variation with redshift of the 
$f_{\text{gas}}$ of quiescent galaxies (hereafter, QGs), which we compared to a collection of 
constraints from the literature. Sect.~\ref{hmm} describes the model, Sect.~\ref{eh?} lists the 
various quiescent and post-starburst (hereafter, ``pSB'') galaxy samples used in this work, while 
we discuss the results of the fit in Sect.~\ref{ahh} and present our conclusions in Sect.~\ref{bye}. 
We assume a \citet{Sal55} initial mass function (IMF) throughout and a $\Lambda$CDM cosmology 
with $H_0=70$\,km\,s$^{-1}$\,Mpc$^{-1}$, $\Omega_{\text{M}}=0.3$, and $\Omega_{\Lambda}=0.7$.

\section{\label{hmm}Model description}

To understand the evolution of the ISM of QGs, we first considered their stellar 
mass function (MF) $\Phi(M_{\star},z)$~and its evolution from $z>3$~to $z\sim0$. Here we
used the \citet{Dvd17} MFs, as they currently cover the largest redshift range; 
for passive galaxies, they are fit up to $z=4$, in redshift bins of 0.5 at $z>1.5$~and 0.3 below. 
The rate at which QGs of a given stellar mass $M_{\star}$~appear is then given by the derivative 
of the MF, $\partial\Phi(M_{\star},z)/\partial{z}$, which for the purposes of this study we forced to always be either 
positive or null; that is to say, we neglected rejuvenation in QGs and assumed that their number density 
can only decrease due to mergers. 
We computed this numerically, interpolating the MFs to an arbitrarily fine redshift grid for 
practicality. However, since information on the evolution of the MFs is, in any case, limited by their 
redshift binning, we did not smooth them to avoid risking inducing spurious behavior. 
The mean or median observable $\langle\xi\rangle$ of a QG population can then be modeled as the 
average of $\xi$ as a function of time, from the beginning of QGs to the epoch of observation, weighted 
by the QG production rate. At fixed mass $M_{\star}$ and redshift $z$, 
\begin{equation}\label{eq:avg}
\begin{aligned}
&\langle\xi(M_{\star},z)\rangle = \frac{\int_{z_{max}}^{z}{\xi\frac{\partial\Phi}{\partial{z'}}
\mathrm{d}z'}}{\int_{z_{max}}^{z}{\frac{\partial\Phi}{\partial{z'}}\mathrm{d}z'}}~~\text{(weighted mean) or}\\
&\langle\xi(M_{\star},z)\rangle = \xi\left(M_{\star},z^* \middle| {\frac{\int_{z_{max}}^{z^*}
{\frac{\partial\Phi}{\partial{z'}}\mathrm{d}z'}}{\int_{z_{max}}^{z}
{\frac{\partial\Phi}{\partial{z'}}\mathrm{d}z'}}=\frac{1}{2}}\right)~~\text{(weighted median),}
\end{aligned}
\end{equation}
\noindent
depending on the type of data to which the model is compared. Here $z_{\text{max}}$~is the epoch 
at which the first QGs appear. Given the redshift limit and binning of the MFs, we adopted 
$z_{\text{max}}=3.5$~\\

\begin{figure}
\centering
\includegraphics[width=0.49\textwidth]{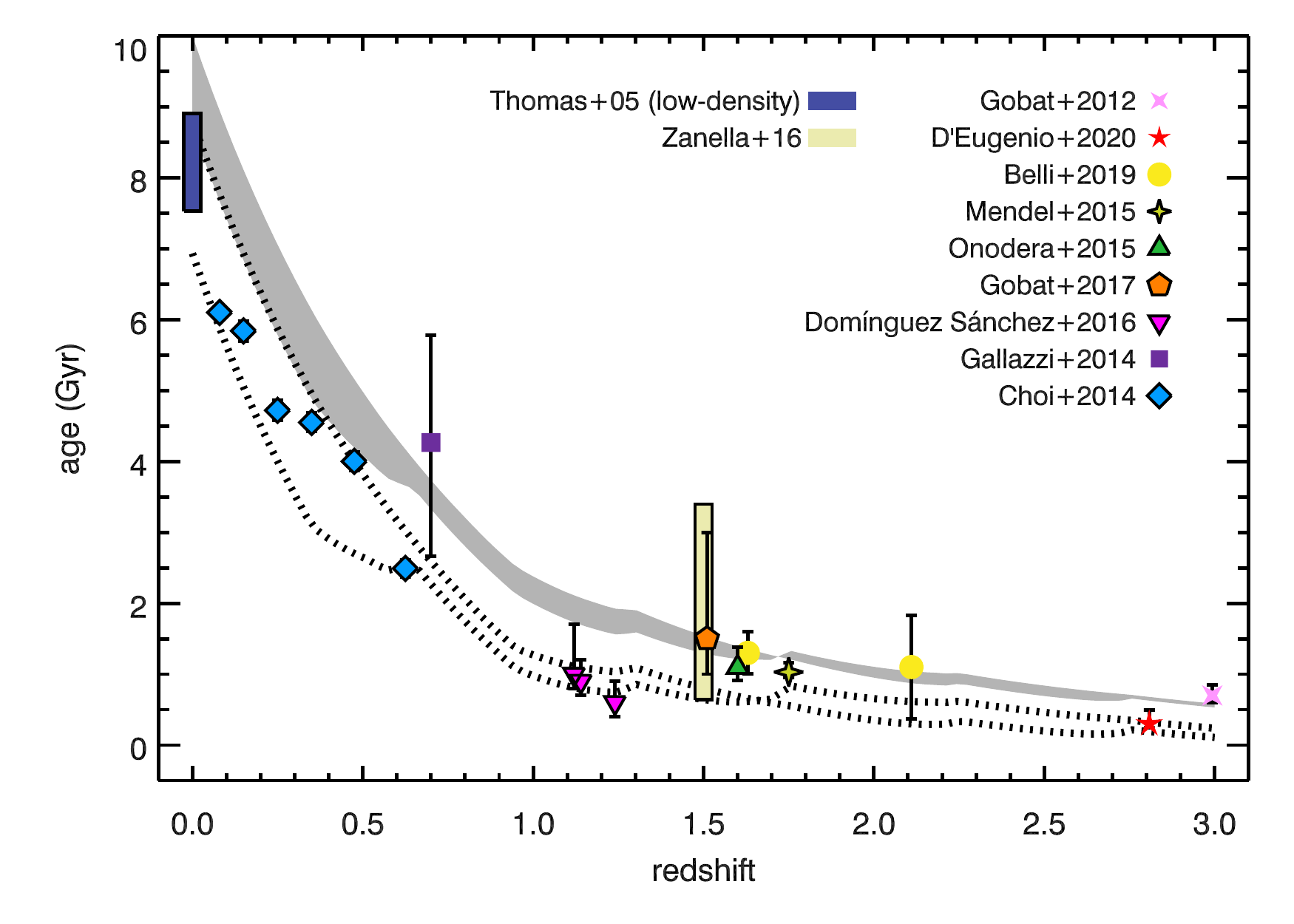}
\caption{Average mass-weighted age and time since quenching (gray and dotted regions, respectively) 
of QGs with $\log(M/M_{\odot})=10.5-11.5$~predicted by the QG formation rate, compared to various field 
samples of $\sim$10$^{11}$\,M$_{\odot}$ QGs. Black-rimmed symbols indicate luminosity-weighted ages, 
while un-rimmed ones show mass-weighted ages. The dark blue and cream rectangles show the age spread 
predicted from the empirical $\text{age}-M_{\star}$~relations of \citet{Tho05} and \citet{Zan16}, 
respectively.
}
\label{fig:etgages}
\end{figure}

For example, Fig.~\ref{fig:etgages} shows the mass- and luminosity-weighted ages of several 
spectroscopic field QGs samples 
\citep{Gal14,Cho14,Men15,Ono15,Dom16,Gob17,Bel19,DEu20}, as well as an individual $z\sim3$~QG 
\citep{Gob12}, compared to the average quenching age (i.e., the look-back time to the quenching 
event) and mass-weighted age predicted by the QG formation rate, two quantities which should 
bracket the luminosity-weighted age. To compute mass-weighted ages, we used a star formation 
history (SFH) starting at $z=10$~and following the main sequence of star formation 
\citep[MS, as parameterized by][]{Sar14} until truncation 
\citep[see, e.g.,][for a similar approach]{Men15}. The predicted ages follow the observed 
trend, reproducing the dynamic range of ages at $z\sim0$~reasonably well \citep{Tho05}, but 
somewhat under-predicting their spread at $z>1$~\citep[e.g.,][]{Bed13,Zan16}. However, we note 
that ages derived from observations using non-MS SFHs, such as exponentially-declining ones, 
can be significantly higher than those assuming a truncated MS SFH. Conversely, high-redshift 
massive QGs might have followed super-MS paths prior to quenching \citep[e.g.,][]{Val20}.\\

In this analysis, we assumed that the cessation of star formation in galaxies can happen through 
either a ``slow'' ($s$) or a ''fast' ($f$) channel, the latter of which we identified with major 
dissipative mergers \citep[e.g.,][]{Scha14}. However, aside from the existence of these two channels, 
we did not posit any specific quenching mechanisms nor, for example, whether the transition from 
star formation to quiescence happened suddenly or in a more gradual way \citep[e.g.,][]{Gut17}. 
The model's second main assumption is that all newly-quenched galaxies start their quiescent 
evolution with the same initial gas fraction, $f_{g,0,s}$~or $f_{g,0,f}$,~depending on their 
quenching channel. We assumed two hypothetical cases for this initial gas fraction: In the first 
``constant'' case, $f_{g,0}$~does not depend on redshift; whereas in the second ``fractional'' case, 
it is a fixed percentage of the evolving molecular gas fraction of the MS 
$f_{g,\text{MS}}(M_{\star},z)$. In this second case, we adopted the $f_{g,\text{MS}}$ 
parameterization of \citet[hereafter S14]{Sar14}.  
Thirdly, we let the remaining gas mass $M_g$~be depleted (either consumed or expelled), without
replenishment at a rate $\Psi$~inversely proportional to a time $t_{\text{dep}}$. The gas mass of 
a QG at time $t$~after quenching is therefore
\begin{equation}\label{eq:quench}
\begin{aligned}
M_{\text{g}}(t) &= M_{\text{g}}(0) - M_{\text{g}}^{-}(t)~~\text{with}~~M_{\text{g}}^{-}(t) = 
\int_{0}^t\Psi(t')\mathrm{d}t'\\
\Psi(t) &= M_{\text{g}}(t)/t_{\text{dep}}\text{,}
\end{aligned}
\end{equation}
\noindent
which we solved numerically in time steps. The initial gas mass immediately after quenching is defined 
as $M_{\text{g}}(0)=M_{\star}(0)f_{g,0}/(1-f_{g,0})$, while $M_{\text{g}}^{-}(t)$ is the gas mass lost 
within time $t$ due an unspecified process with rate $\Psi$. The gas fraction at time $t$~after 
quenching is then $f_{\text{gas}}(t)=M_g(t)/(M_{\star}(0)+M_{\star}^{+}(t)+M_g(t))$, with 
$M_{\star}^{+}(t)$ being the change in stellar mass within time $t$. If gas depletion is due to star 
formation, $M_{\star}^{+}(t) = M_{\text{g}}^{-}(t) - M_{\star}^{-}(t)$, where $M_{\star}^{-}(t) < 0$ 
accounts for stellar death and depends on past star formation and the IMF. If gas is depleted by other 
processes, $M_{\star}^{+}(t)= M_{\star}^{-}(t)$. Here we chose to keep $M_{\star}^{+}(t)=0$, both for 
simplicity and so as to not make assumptions regarding the physical processes responsible for the gradual 
depletion of the remaining gas and dust in QGs. Furthermore, it is not known how much of the gas in 
high-redshift galaxies is in the molecular phase.
In addition, whether or not $\Psi$~arises from star formation makes little to no difference at our current 
level of precision, with either alternative yielding very similar $f_{\text{gas}}(t)$~values 
(see Sect.~\ref{ahh}).\\

In the case of pSB galaxies, which are not age-composite populations, we only considered the 
aforementioned gas depletion, starting (somewhat arbitrarily) within $t\leq500$\,Myr of the epoch of 
observation and assuming an initial gas fraction of $f_{g,0,s}$. For true QGs, on the other hand, gas 
depletion must be computed for every possible epoch from $z_{\text{max}}$ to the redshift of observation, 
for both $f_{g,0,s}$ and $f_{g,0,f}$, and combined with both the QG formation rate and major merger rate 
per descendant galaxy. To parameterize the latter, we used the fitting function of \citet{Rod16}:
\begin{equation}\label{eq:mergers}
\frac{\mathrm{d}n_{\text{merg}}}{\mathrm{d}\mu\mathrm{d}t} = 
A(z)\left(\frac{M_{\star}}{10^{10}\,M_{\odot}}\right)^{\alpha(z)}\left[1+\left(\frac{M_{\star}}
{M_0}\right)^{\delta(z)}\right]\mu^{\beta(z)+\gamma\,\log(\frac{M_{\star}}{10^{10}\,M_{\odot}})}
\text{,}
\end{equation}
\noindent
where $\mu$~is the mass ratio, for which we assumed that a cut-off value of $\mu_{\text{min}}=1/3$, 
$M_{\star}$~is the descendant mass, $M_0=2\times10^{11}\,M_{\odot}$, and the redshift dependence 
of parameters $\alpha$, $\beta$, and $\delta$ is of the kind $X(z)=x_0(1+z)^{x_1}$, using the values 
of $x_i$~given in that paper.
The $\langle f_{\text{gas}}\rangle$~of a population of QGs with stellar mass $M_{\star}$~at 
redshift $z$~is then the weighted average (mean or median, depending on the data to which it is 
compared) of
\begin{equation}\label{eq:fgas}
\begin{aligned}
&\xi =f_{\text{gas},s}(t_z-t_{z'})(1-w_{z'}) + f_{\text{gas},f}(t_z-t_{z'})w_{z'}~~\text{, where}\\
&w_{z'} = \int_{\mu_{\text{min}}}^{1}\frac{\mathrm{d}n_{\text{merg}}}{\mathrm{d}\mu\mathrm{d}t}
\frac{\mathrm{d}t}{\mathrm{d}z'}\mathrm{d}\mu~~\text{,}
\end{aligned}
\end{equation}
\noindent
between $z'=z$~and $z'=z_{\text{max}}$, as defined in Eq.~\ref{eq:avg}. A schematic representation 
of the model is shown in Appendix~\ref{appendix:model}.\\

\section{\label{eh?}Gas samples}

As in G18, we compiled existing constraints on the molecular gas fraction $f_{\text{gas}}$~of 
quiescent and pSB galaxies from recent literature, namely: local QGs consisting of the 
\textsc{ATLAS}$^{\textsc{3D}}$  \citep{You11,Cap13,Dav14} and \textsc{HRS} \citep{Bos14,Lia16} 
ETG samples as well as the samples of pSB
galaxies (hereafter, the 
``low-z pSB'' sample) of \citet{Fre15} and \citet{Ala16}; at low and intermediate redshift, the 
ETG sample of \citet{Spi18}  and the pSB sample of \citet{Sue17}; at intermediate and high redshift, 
constraints from \citet{Hay18} on gas in $z\sim1.46$~cluster ETGs, as well as on individual galaxies 
from \citet{Sar15}, \citet{Bez19}, and \citet{Rud17}. Given its size, we divided the 
\textsc{ATLAS}$^{\textsc{3D}}$~sample into high- and low-mass subsamples, choosing 
$5\times10^{10}$\,M$_{\odot}$~as the cut-off mass.
In addition, we also included $f_{\text{gas}}$~estimates derived from the (median) stacked FIR 
spectral energy distributions of ETGs at $z\sim1.8$~(G18; 977 galaxies), $z\sim1.2$, $z\sim0.8$, 
and $z\sim0.5$ 
(1394, 1536, and 563 galaxies, respectively; Magdis et al., submitted, hereafter M20). 
Finally, at higher redshift ($z\sim3$), we converted star formation rates (SFR) estimated from 
spectroscopy \citep{Sch18c,DEu20} into gas masses assuming the star formation efficiency found 
by G18. 
As a consequence of our $z_{\text{max}}=3.5$, we did not include higher-redshift quiescent 
galaxies \citep{Gla17,Sch18b,Tan19,Val20} in the analysis and considered $z\sim3$ galaxies as pSB.
The dust-based estimates of G18 and M20 (and, by extension, the $z\sim3$~semi-constraints) 
assume a gas-to-dust ratio (G/D). It is dependent on metallicity, which is presumed to be 
solar or higher owing to both the relatively high gas-phase metallicity of MS galaxies at 
$z\lesssim1$~\citep[e.g.,][]{Mnu10} and the already high stellar metallicities of QGs at 
$z>1$~\citep{Ono15,Est19}. Here we adopted an intermediate value between the solar and supersolar 
G/Ds used in M20, and we increased the error bars of these points to include both the solar and 
supersolar confidence estimates.
These various samples, which are summarized with their selection criteria in Table~\ref{tab:samples}, 
combine into a nonhomogeneous dataset: some were specifically selected as ETGs, and others were 
based on varying degrees of quiescence. In particular, pSB galaxies are not necessarily truly 
quiescent and could, in principle, resume normal star formation. However, as a possible 
precursor of QGs, they provide useful, though not constraining (see Sect.~\ref{ahh}), comparison 
samples for the model. Here we refer to all equally as either QGs or pSB galaxies, and we  make the 
assumption that, on average, these different samples are not otherwise significantly biased with 
regard to their gas content compared to the full population, given each mass limit and type.\\

Model gas fractions were then computed, as described in Sect.~\ref{hmm}, for each sample according 
to its median mass and redshift. All samples and subsamples were fit together with no distinction 
for their median mass, as there is currently not enough statistics for a mass-specific fit. 
We let $f_{g,0,s}$~and $f_{g,0,f}$~vary from 0 to 0.5 -- that is to say about the MS value at $z\sim2$ -- 
in steps of $\Delta f_{g,0}=0.01$. We also allowed $t_{\text{dep}}$~to vary from 0.1 to 8\,Gyr with 
$\Delta t_{\text{dep}}=100$\,Myr\footnote{The model grid is publicly available at 
\url{www.georgiosmagdis.com/software/}}. These values were then fit, via $\chi^2$~minimization, to 
the median $f_{\text{gas}}$~and scatter of each sample to mitigate the effect of outliers. In each 
case, the residuals were multiplied by the number of objects in the sample. Samples which contain 
both detections and non-detections were each split in two subsamples of detected and 
non-detected objects, respectively, using 3$\sigma$~upper limits for the latter. In this case, and 
for galaxies explicitly selected or identified as pSB (i.e., with estimated ages $\leq500$\,Myr), we 
adopted a data censoring approach where the model could take any value of $f_{\text{gas}}(t)$ within 
the 3$\sigma$~limit and the $t\leq500\,\text{Myr}$ range, respectively.\\

\section{\label{ahh}Results and discussion}

\begin{figure}
\centering
\includegraphics[width=0.49\textwidth]{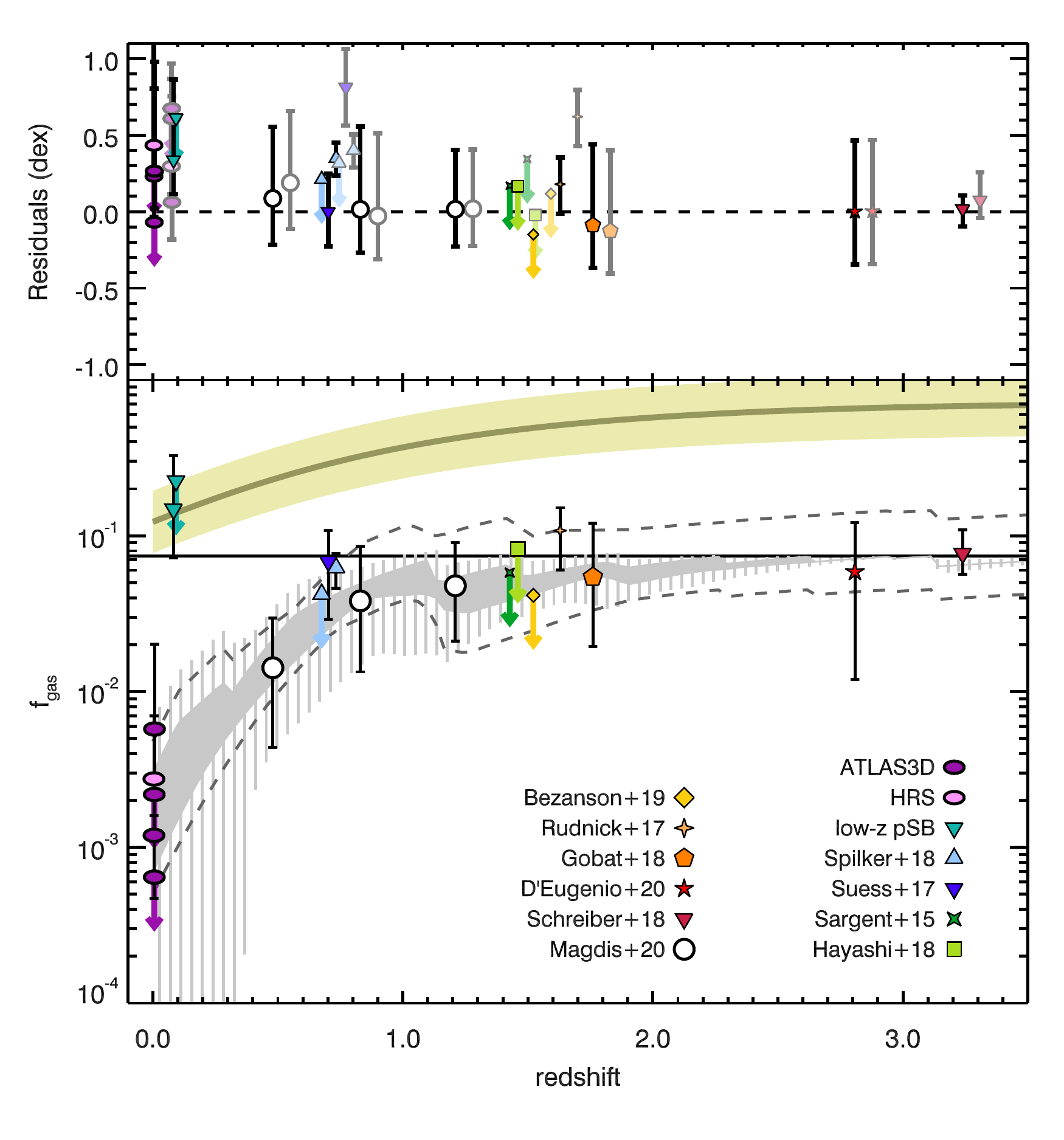}
\caption{\emph{Bottom panel}: Evolution of the $\langle f_{\text{gas}}\rangle$ of quiescent 
galaxies as a function of redshift. The various symbols with error bars or upper-limit arrows 
show the median $f_{\text{gas}}$~of the QG and pSB samples. The solid and hatched light gray 
regions trace the envelope of the best-fit models with constant $f_{g,0}$, 
within the range of median stellar masses of the samples, and its predicted 1$\sigma$~scatter, 
respectively. The dark dashed gray lines show the best-fit envelope for the fractional case model 
and the horizontal black line shows the best-fit $f_{g,0,s}$~for the constant case model.
For comparison, the evolution of $f_{\text{gas}}$~of 5$\times$10$^{10}$\,M$_{\odot}$~MS 
galaxies, as parameterized by S14, is shown as a tan line and its scatter envelope is shown 
as a solid cream region.
\emph{Top panel}: Median residuals of the fits with constant and fractional 
$f_{g,0}$~(solid and clear symbols, respectively, the latter having been offset for 
clarity) for the same samples. The error bars show the dispersion of residuals. Smaller symbols 
\citep{Sar15,Rud17,Bez19} indicate single objects.}
\label{fig:fgevol}
\end{figure}

\begin{figure}
\centering
\includegraphics[width=0.49\textwidth]{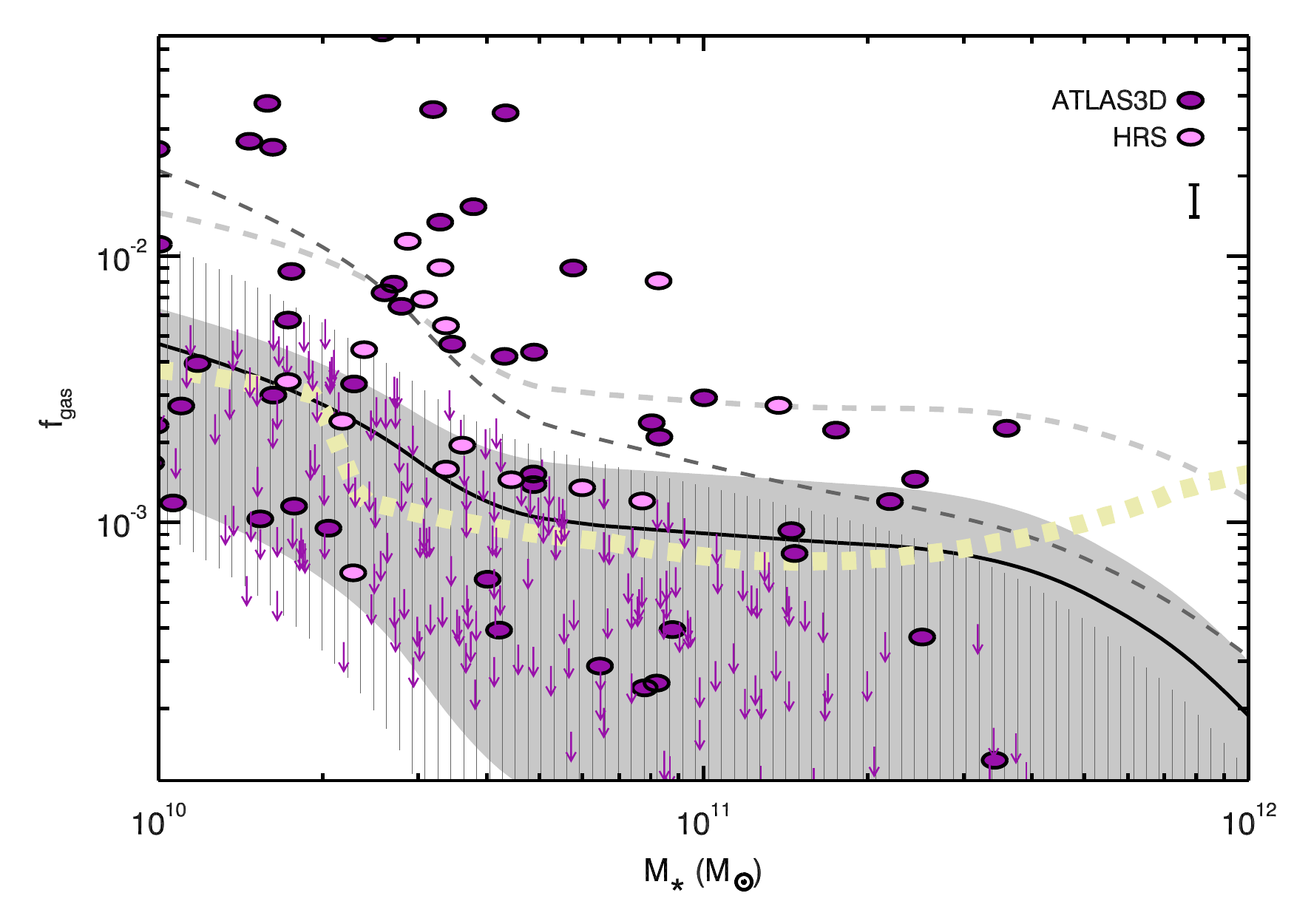}
\caption{Gas fraction of $z\sim0$~ETGs in the \textsc{ATLAS}$^{\textsc{3D}}$ and \textsc{HRS} 
samples, as a function of stellar mass. Colored ellipses indicate detections 
(a typical error bar is shown in the upper right corner), and downward-pointing arrows indicate the 
3$\sigma$~upper limits of non-detections. The solid light-gray and hatched dark-gray regions 
correspond to constant and fractional-case models with $f_{g,0,f}=0$, respectively, within 
the 1$\sigma$ confidence interval of their respective fits. The dashed lines show the 
1$\sigma$~scatter in $f_{\text{gas}}$~predicted by the models, with respect to the best fit, 
shown here as a solid line. On the other hand, the best-fit model for the constant 
case and varying $f_{g,0,f}$ is shown as a thick cream dashed line.}
\label{fig:fgz0}
\end{figure}

Both the constant and fractional models reproduce the apparent evolution of existing 
constraints within their uncertainties (Fig.~\ref{fig:fgevol}).
At $z\gtrsim1$, the evolution of $\langle f_{\text{gas}}\rangle$~is mostly driven by the QG 
formation rate and the gas fraction is kept relatively high by the steady emergence of newly 
quenched galaxies (see Appendix~\ref{appendix:qrate}). Some fluctuations are present, which 
are a consequence of the redshift binning of the MFs used to compute the model. 
At lower redshift, the high-mass end of the QG MF changes only slightly and the 
remaining gas gets steadily depleted. As a consequence of this difference in the formation rate 
between high- and lower-mass QGs, their respective $\langle f_{\text{gas}}\rangle$~tracks start 
diverging which, at $z\sim0$, causes an anticorrelation between the gas and stellar masses of 
QGs, as seen in the data (Fig.~\ref{fig:fgz0}).\\

The model fit is relatively robust with respect to the samples used, with 
$\langle f_{\text{gas}}\rangle\sim0.1$ varying by less than 0.3\,dex when only fitting part 
of the sample (e.g., only local galaxies or only intermediate-to-high redshift ones; see 
Appendix~\ref{appendix:subfits}). We also find that choosing different MFs does not appreciably 
affect the best-fit parameters 
\citep[e.g., ][who derived them from a larger but shallower survey area]{Ilb13}. In all cases, 
the resulting models stay well within the uncertainties of the data.
It is therefore interesting to note that the putative gas fractions of $z\sim3$~ETGs, 
which we have speculatively estimated here from emission-line SFRs, appear to be 
consistent with the predictions of the model. This suggests in turn that the star 
formation efficiency of quiescent galaxies might not have changed much between 
$z\sim3$~and $z\sim1.8$.
\\

\begin{table}
\caption{Best-fit parameters}
\centering
\renewcommand{\arraystretch}{1.5}
\setlength{\tabcolsep}{10pt}
\begin{tabular}{c c c c}
\hline\hline
Case & $f_{g,0,s}$ & $f_{g,0,f}$ & $t_{\text{dep}}$\,(Gyr)\\
\hline
constant & $0.17^{+0.6}_{-0.6}$ & 0$^{*}$ & $2.2^{+0.6}_{-0.9}$\\
& $0.07^{+0.06}_{-0.01}$ & $0.07^{+0.02}_{-0.01}$ & $1.7^{+0.1}_{-0.4}$\\
\hline
& $f_{g,0,s}/f_{g,\text{MS}}$ & $f_{g,0,f}/f_{g,\text{MS}}$ & $t_{\text{dep}}$\,(Gyr)\\
\hline
fractional & $0.12^{+0.07}_{-0.06}$ & 0$^{*}$ & $4.0^{+3.5}_{-2.7}$\\
& $0.05^{+0.07}_{-0.01}$ & $0.05^{+0.07}_{-0.02}$ & $2.9^{+1.0}_{-1.1}$\\
\hline
\end{tabular}
\justify
$^{*}$~fixed parameter.
\label{tab:params}
\end{table}

Table~\ref{tab:params} shows the best-fit parameters in the constant and fractional 
cases for two different instances: one where the initial gas fraction of fast-quenched 
QGs was fixed to $f_{g,0,f}=0$~and another where it was let to vary freely. 
Both models fall well within the uncertainties of the constraints, despite being based 
on different assumptions. This is likely a consequence of the evolution of $f_{g,\text{MS}}$ 
being rather shallow, compared to the trend seen in QGs, in particular at $z>1$~where the 
majority of massive QGs formed. However, while the constant case is also consistent with 
the relatively large gas fractions seen in local pSB galaxies \citep{Fre15,Ala16}, the 
fractional model underpredicts them by a factor of $\sim$10.
The depletion time suggested by the former is also similar to that estimated for 
intermediate-age pSB galaxies \citep{Row15} and intermediate- to high-redshift ETGs (G18, 
M20), while being slightly higher than that of local gas-rich ETGs \citep{You11,Dav14}. 
We note that, given the relatively low $f_{g,0}$~of QGs, their stellar mass at $z\sim0$ 
would not change much whether the gas were fully converted into stars or exhausted through 
other processes. Moreover, assuming that gas depletion is due to star formation would not 
change the fit or its parameters significantly (see Fig.~\ref{fig:subfits}) since the 
depletion time is constrained by the evolution of $f_{\text{gas}}$~at $z<1$, while 
the ratio between $f_{g,0}$~and $t_{\text{dep}}$~is constrained by the overall data.
At the current level of precision of the data, the model is therefore not reliant on a 
specific depletion mechanism and our results do not change if gas, rather than being consumed, 
is expelled from galaxies on the same timescale for example. Indeed, the $t_{\text{dep}}$~we 
find for the constant case is comparable to the timescale of dust removal found in low-redshift 
ETGs \citep{Mic19}, a process that does not necessarily involve star formation.\\

On the other hand, local massive MS galaxies tend to have $f_{\text{gas}}\sim 0.1-0.2$,~while 
not being ostensibly on the cusp of becoming quiescent \citep[e.g.,][]{Ler09,Ler13}, 
hence a gas fraction of $\leq$$f_{g,0,s}$ is not likely to be a sufficient condition for quenching. 
The cessation of star formation in a galaxy can clearly be mediated by a variety of mechanisms, 
some of which might not be adequately represented by either of the quenching channels that this 
intentionally simple model considers (e.g., environmentally-driven ones). 
Additionally, we did not account for post-quenching mergers, which can either somewhat increase 
a QG's mass while keeping it quiescent or rejuvenate it via gas reaccretion, corresponding to dry 
and dry-wet mergers, respectively. At the end of each episode, the resulting object would be treated 
as newly quenched by our model. However, its gas content would likely be substantially different 
from that of pSB galaxies. In particular, we should expect dry merger remnants to be even more 
gas-depleted than noninteracting QGs \citep[see, e.g.,][]{Bas17}; while in its present state, 
our model would assign a higher $f_{\text{gas}}$~to them. This is illustrated by the different 
behavior at $>3\times10^{11}$\,M$_{\odot}$ of tracks with fixed and varying $f_{g,0,f}$ 
(Fig.~\ref{fig:fgz0}). Finally, even rejuvenated QGs are typically identified as sub-MS 
\citep{Cha19,Mnc19}. Model predictions might therefore substantially deviate from observations 
for galaxy populations where these processes play a significant role.\\

\section{\label{bye}Conclusions}

We have constructed a simple model to interpret the apparent evolution of the ISM in QGs. 
It assumes that galaxies can quench through two different channels, one of which 
is related to major merging events, and that no ISM replenishment takes place after 
quenching. On the other hand, we make no further suppositions otherwise as to the physical 
processes involved in either quenching the galaxies or keeping them quiescent. A comparison with 
recent constraints suggests that quiescent galaxies begin their passive evolution with 
non-negligible amounts of gas left, $f_{\text{gas},0}\sim0.1$, which is progressively depleted 
with a timescale of $1.5-2.5$\,Gyr. At early times, the average gas fraction of the QG population 
is kept relatively high by newly quenched galaxies. Below $z\sim1$, on the other hand, the 
formation rate of massive QGs drops dramatically and, as this population grows older, its 
average $f_{\text{gas}}$ decreases slowly.\\

However, conclusively distinguishing between different possible cases for the initial gas 
fraction would require more complete datasets than are currently available. Likewise, a 
more sophisticated treatment of quenching channels and their associated consumption might 
be warranted once the quiescent galaxy population has been more finely sampled in redshift, 
mass, and environment. 
Nevertheless, this present analysis shows that the emerging trends in the evolution of 
gas in quiescent galaxies can already be broadly explained using a minimal number of 
assumptions. This suggests that, as with star forming galaxies, the mechanisms regulating 
the ISM content of quiescent galaxies operate consistently over most of the history of the 
Universe.

\begin{acknowledgements}

We thank I. Davidzon for fruitful discussions and the anonymous referee for their help in 
improving this letter. GEM and FV acknowledge the Villum Fonden research grant 13160 
``Gas to stars, stars to dust: tracing star formation across cosmic time'' and the Cosmic 
Dawn Center of Excellence funded by the Danish National Research Foundation under then 
grant No. 140. FV acknowledges support from the Carlsberg Foundation research grant 
CF18-0388 ``Galaxies: Rise And Death''.

\end{acknowledgements}

\begin{appendix}

\section{\label{appendix:model}Scheme of the model}

Here we show a schematic representation of our model, with its two conditional paths. 
For pSB samples, we assume that quenching occurred within the last 500\,Myr and allow for 
any gas fraction predicted by Eq.~\ref{eq:quench} within that time; on the other hand, the 
$f_{\text{gas}}$ of a general QG population is a composite weighted by the QG formation 
rate and the major merger rate.

\tikzstyle{block1}=[rectangle,draw,rounded corners,text centered,text width=6em]
\tikzstyle{block2}=[rectangle,draw,rounded corners,text centered,text width=8em]
\tikzstyle{line}=[draw,-latex']
\tikzstyle{line2}=[draw,-latex']
\begin{center}
\begin{tikzpicture}[auto, node distance=90pt]
\node [block1] (gasdep) {Gas depletion Eq.~\ref{eq:quench}};
\node [block1, below of=gasdep] (psb) {Post-starburst galaxy};
\node [block2, right of=gasdep] (qrate) {QG formation rate\\$\frac{\partial\Phi(M_{\star},z)}{\partial{z}}$};
\node [block1, right of=qrate] (mrate) {Merger rate\\Eq.~\ref{eq:mergers}};
\node [block2, right of=psb] (QG) {Composite QG population\\Eq.~\ref{eq:fgas}};
\begin{scope}[node distance=50pt]
\node [block2, below of=QG] (avg) {Average $f_{\text{gas}}$\\Eq.~\ref{eq:avg}};
\end{scope}
\path [line] (gasdep) -- node[above, near end, fill=white]{$t\leq500$\,Myr} (psb);
\path [line] (gasdep) -- (QG);
\path [line] (qrate) -- (QG);
\path [line] (mrate) -- (QG);
\path [line] (QG) -- (avg);
\end{tikzpicture}
\end{center}

\section{\label{appendix:samples}Quiescent galaxy samples}

In Table~\ref{tab:samples} we list the various QG and pSB samples used in this analysis. We 
include the following information: the number of objects in each sample or subsample (N), 
its redshift range, median stellar mass, median $f_{\text{gas}}$, the type of objects 
it contains (whether quiescent or post-starburst), and the principal selection criterion. 
For the latter, we consider selection on the basis of early-type morphology (``morphology''), 
absorption or emission features in their spectra (``spectrum''), their distance to the MS (``sSFR''), 
or color-color criteria (``color''). However, this does not include additional refinements to 
the selection criterion, such as mid-IR non-detection for color-selected QGs (e.g., G18 and M20). 
As described in Sect.~\ref{eh?}, samples have been split into detections and 3$\sigma$~upper 
limits when appropriate.

\begin{table*}
\caption{Quiescent and post-starburst galaxy samples.}
\centering
\renewcommand{\arraystretch}{1.5}
\setlength{\tabcolsep}{10pt}
\begin{tabular}{c c c c c c c}
\hline\hline
Sample & N & $z$ range & $\langle \log M_{\star}\rangle$ & $\langle \log f_{\text{gas}}\rangle$ 
& Type & Selection\\
\hline
\textsc{ATLAS}$^{\textsc{3D}}$ high-$M_{\star}$ & 16 & $0.0034-0.0077$ & 11.16 & -2.92 & QG & morphology\\
& 86 & $0.024-0.0095$ & 10.98 & $<-3.67$ & QG & morphology\\
\textsc{ATLAS}$^{\textsc{3D}}$ low-$M_{\star}$ & 37 & $0.0027-0.096$ & 10.24 & -2.24 & QG & morphology\\
& 115 & $0.0025-0.0107$ & 10.39 & $<-3.14$ & QG & morphology\\
\textsc{HRS} & 15 & $0.0027-0.027$ & 10.53 & -2.56 & QG & morphology\\
low-$z$ pSB & 55 & $0.013-0.197$ & 10.71 & -0.76 & pSB & spectrum\\
& 9 & $0.038-0.166$ & 10.67 & $<-1.09$ & pSB & spectrum\\
\citet{Spi18} & 4 & $0.60-0.75$ & 11.20 & -1.18 & QG & sSFR\\
& 5 & $0.65-0.71$ & 11.31 & $<-1.85$ & QG & sSFR\\
\citet{Sue17} & 2 & $0.66-0.75$ & 11.45 & -1.13 & pSB & spectrum\\
M20 low-$z$ & 563 & $0.30-0.65$ & 11.17 & -1.70 & QG & color \\
M20 intermediate-$z$ & 1536 & $0.65-1.00$ & 11.20 & -1.23 & QG & color\\
M20 high-$z$ & 1394 & $1.0-1.4$ & 11.16 & -1.14 & QG & color\\
\citet{Sar15} & 1 & 1.43 & 11.82 & $<-1.71$ & QG & color\\
\citet{Hay18} & 12 & 1.46 & 11.04 & $<-1.55$ & QG & color\\
\citet{Bez19} & 1 & 1.52 & 11.42 & $<-1.85$ & QG & spectrum\\
\citet{Rud17} & 1 & 1.63 & 11.44 & -0.92 & pSB & sSFR\\
G18 & 997 & $1.5-2.2$ & 11.04 & -1.08 & QG & color\\
\citet{Sch18c} & 4 & $3.22-3.72$ & 11.28 & -1.08 & QG/pSB & color\\
\citet{DEu20} & 9 & $2.39-3.23$ & 11.23 & -1.21 & QG/pSB & color\\
\hline
\end{tabular}
\justify
\label{tab:samples}
\end{table*}

\section{\label{appendix:qrate}Quiescent galaxy formation rate}

\begin{figure}
\centering
\includegraphics[width=0.49\textwidth]{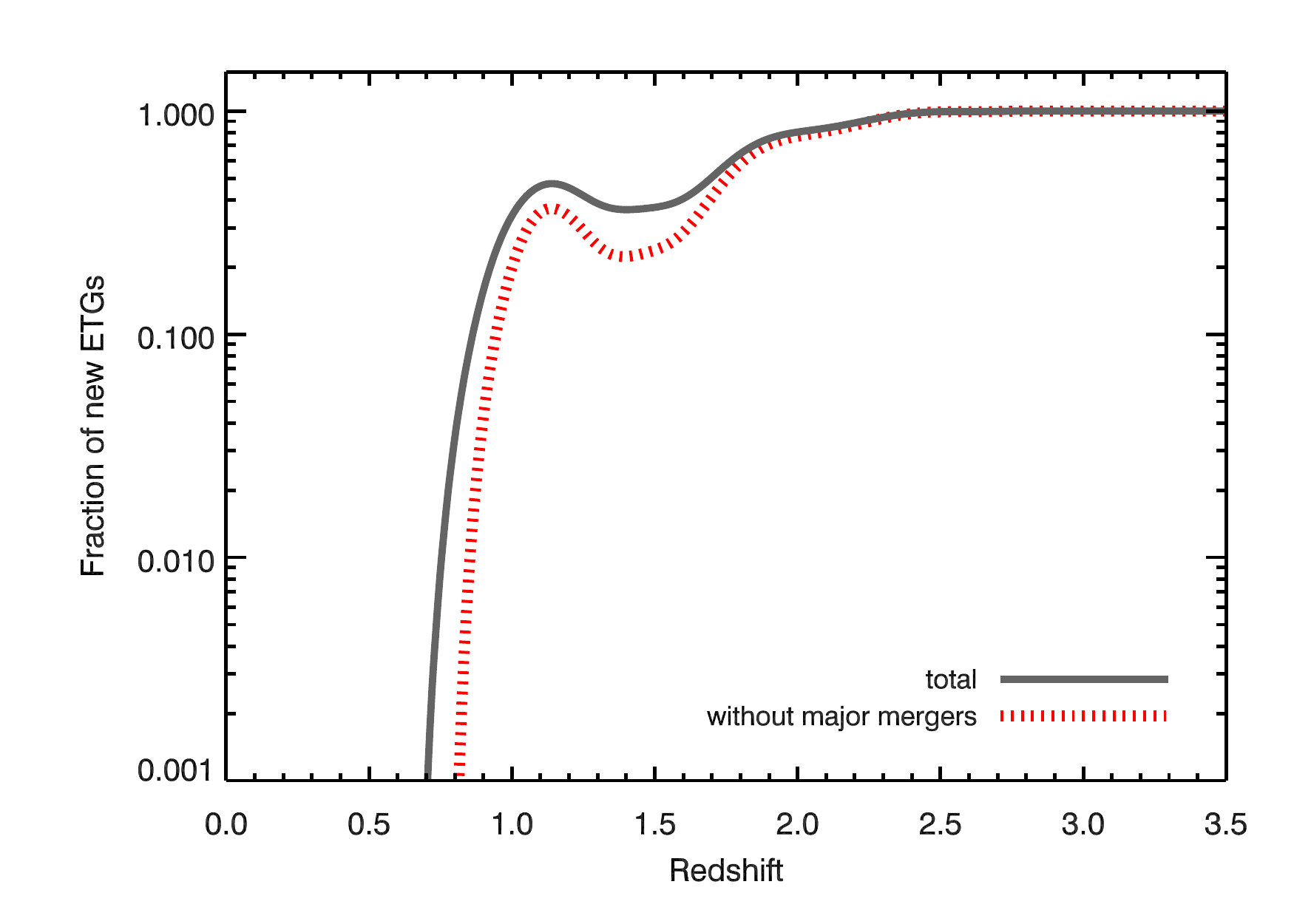}
\caption{Smoothed evolution of the fraction of $\leq500$\,Myr 
$2\times10^{11}$\,M$_{\odot}$ QGs as a function of redshift (dark curve). 
The red dotted curve shows the pSB fraction after subtracting the descendants 
of major mergers (i.e., with mass ratios $\mu\geq1/3$).}
\label{fig:etgrate}
\end{figure}

In Fig.~\ref{fig:etgrate} we show the fraction of pSB galaxies (i.e., 
galaxies quenched within the last 500\,Myr) predicted by the \citet{Dvd17} 
MFs, as a function of redshift and for a stellar mass of 
$2\times10^{11}$\,M$_{\odot}$. Most of the QG population at $z>1$~is therefore 
expected to be newly-quenched, while at $z\lesssim0.5$~the fraction of 
$2\times10^{11}$\,M$_{\odot}$~pSB is expected to be negligible.

\section{\label{appendix:subfits}Subsample fits}

To test the dependency of our modeling to the various QG and pSB samples, we performed a
series of tests where the model was fit to only part of the full set of constraints. We 
considered the following subsets: one that only contains local QGs (i.e., the 
\textsc{ATLAS}$^{\textsc{3D}}$~and \textsc{HRS} samples); one with only 
intermediate-to-high redshift samples, ignoring the local QG and pSB samples; one which 
excludes pSB galaxies (i.e., which contains only QG samples); one with only ``direct'' 
detections and constraints, excluding the constraints derived from FIR SED fitting (i.e., 
G18, M20, and higher-z QGs). Finally, we also carried out a fit of the full sample assuming 
that gas depletion in QGs is due to star formation.
We find that the model tracks stay within a factor of 2-3 of each other, with the largest deviation 
occurring at $z=0$~between the full-set fit and the nonlocal one, the latter predicting slightly 
higher $f_{\text{gas}}$;~although, it is still consistent with local QG values.

\begin{figure}[!h]
\centering
\includegraphics[width=0.49\textwidth]{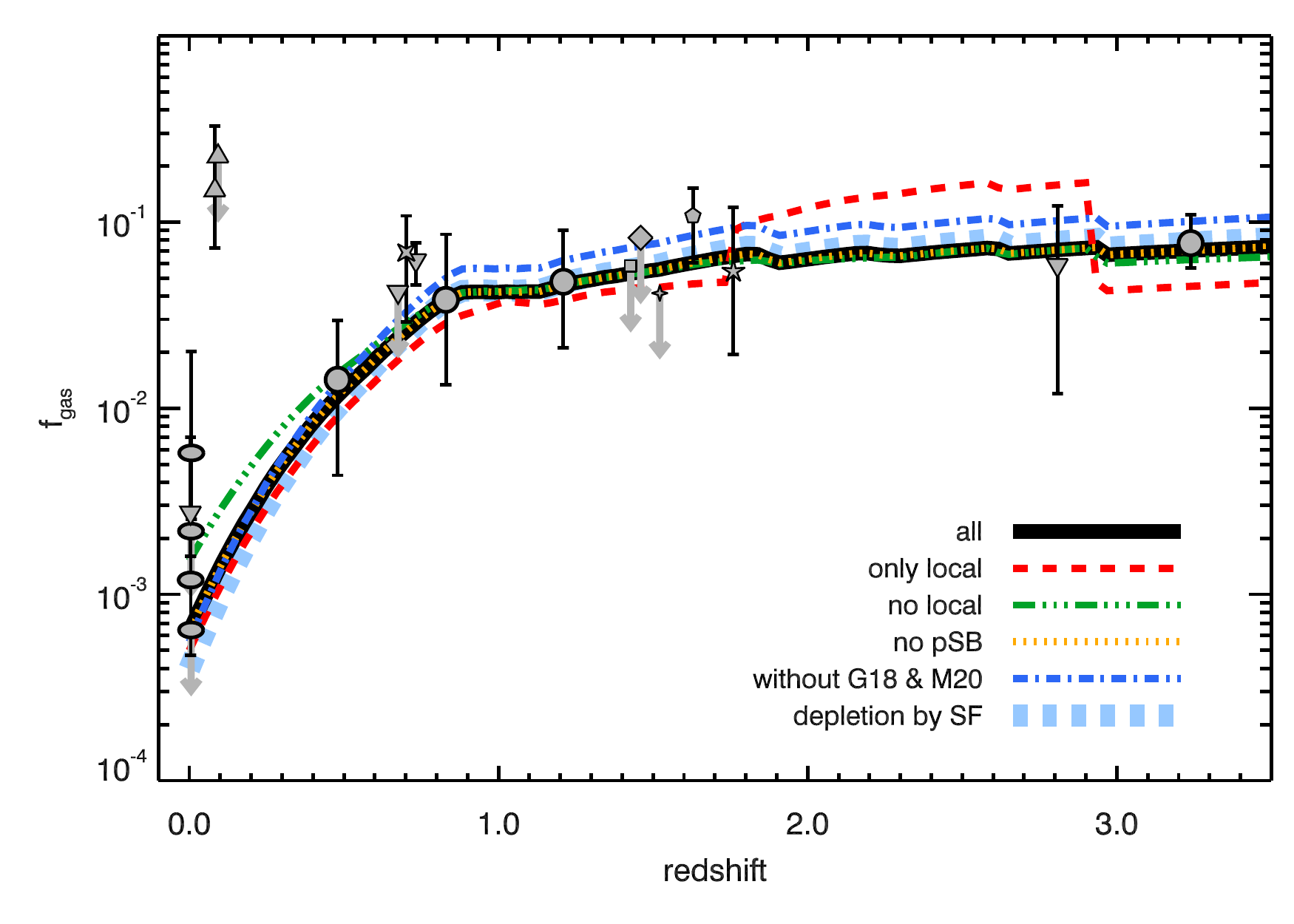}
\caption{$\langle f_{\text{gas}}\rangle$~evolution when fitting all data points, 
as in Fig.~\ref{fig:fgevol} (thick black curve), only local objects (dashed red curve), only 
intermediate-to-high redshift galaxies (green dash-dot-dotted curve), or ignoring either 
pSB galaxies (dotted yellow curve) or dust-derived constraints (G18 and M20; dash-dotted 
blue curve). For comparison, the best-fit model with SF gas depletion is shown as a thick 
dashed light-blue line. The tracks shown here are for a 1.5$\times$10$^{11}$\,M$_{\odot}$~QG, 
while the gray points show the various samples used, as in Fig.~\ref{fig:fgevol}.}
\label{fig:subfits}
\end{figure}

\end{appendix}

\end{document}